\documentstyle[prl,twocolumn,amsfonts,aps]{revtex}

\begin{document}
\input epsf

\title{An Uncontrolled Toy That Can Walk But Cannot Stand Still}

\author{Michael J. Coleman and Andy Ruina }

\address{Department of Theoretical and Applied Mechanics\\
Cornell University, Ithaca, NY\ \ \ 14853-7501 USA \\ 
{\em Submitted to Physical Review Letters,  June 14, 1997} }

\date{revised \today}

\maketitle

\abstract{
We built a simple two-leg toy that can walk stably with no control system. It
walks downhill powered only by  gravity.   It seems to be the first  McGeer-like
passive-dynamic walker that is statically unstable in all standing positions,
yet is  stable in motion.   It is one of few known mechanical devices  that
are  stable  near a statically  unstable configuration  but do not depend on 
spinning parts.  Its design is loosely based on simulations  which do not
predict its observed stability. Its motion highlights the possible role of
uncontrolled  nonholonomic mechanics in  balance. }


\paragraph*{\bf Introduction.}
Human walking on level ground involves dynamic balance which, if viewed in a
course--grained way, is presumably asymptotically stable.  This observed
stability of walking  must depend on some combination of neurological control
and  mechanical  features.  The common view is that neuro-muscular control is
responsible for this balance. To what extent is neuro-muscular coordination of
animal  locomotion, say human walking, really necessary? The bold proposal of
McGeer~\cite{mcgeer7:1989,mcgeer3:1990,mcgeer4:1990,mcgeer6:1992,mcgeer9:1992,mcgeer8:1993}
is that much of the  stabilization of walking might be understood without 
control. 

The possibility that asymptotically stable balance can be achieved without
control is somewhat unintuitive since top-heavy  upright things tend to fall
down when standing still or, more generally, since  dynamical systems often
run away from potential energy maxima. Two mechanics issues that bear on such
stability considerations are that: 1)   Hamiltonian (conservative and
holonomic) dynamical systems cannot have  asymptotic stability, and 2) 
conservative {\em non}holonomic systems can have  asymptotically (exponentially)
stable steady motions  in some variables while at most mildly unstable
in the others, as recalled in Zenkov, {\em et al.}~\cite{zenkov:1997}.

Since before the clever patent of Fallis in 1888~\cite{fallis:1888} (the oldest
reference we have), there have been two and four leg passive-dynamic walking
toys that either walk downhill or that walk on level ground when pulled by a
string.  All such toys that we know about are statically stable when they are
not walking. While their motion is engaging to watch, their dynamic stability
is perhaps not so great a surprise.

\paragraph*{\bf McGeer's passive-dynamic walkers.}
Inspired by a double pendulum  simulation of swinging legs
~\cite{mochon2:1980} and by simple walking toys, McGeer
successfully sought and found two-dimensional, straight-legged and kneed
walking model designs that displayed graceful, stable, human-like walking on a
range of shallow slopes with no actuation (besides gravity) and no control.
McGeer termed the motions of these machines {\em passive-dynamic} walking. 
All of McGeer's successful designs, as well as those of his imitators thus
far~\cite{garcia:1997,coleman:1997,goswami:1997b}, have been more-or-less
constrained against falling over sideways so that their dynamic balance is
fore-aft only.  These  machines cannot stand stably upright except when
their legs
are spread fore and aft. The dynamic  stability of these devices could be
dependent on the static stability of this  spread-leg configuration which is
visited momentarily during walking.

While human walking motion is mostly in the sagittal (fore-aft and
vertical) plane,  the stability of out-of-plane (sideways) motions is an issue
important to a more complete understanding of three-dimensional walking.  
McGeer's~\cite{mcgeer6:1992} numerical 3-D studies only led to  unstable 
periodic motions.  Fowble and Kuo~\cite{kuo:1996} numerically simulated a 
passive-dynamic 3D model of walking  but also did not find stable passive 
motions.

Our recent investigations of walking balance have been based on attempts to 
design  mechanisms that vaguely mimic human geometry and walk without
control. This paper describes one such primitive design (first  reported
in~\cite{coleman3:1997}) which extends to three dimensions, at least 
experimentally, the remarkable two-dimensional walking mechanisms of  McGeer.

\paragraph*{\bf Spinning parts and nonholonomic constraints.} Humans are
notably lacking in gyros, flywheels or other spinning parts. Things with
spinning parts, like tops and gyros, are well known to be capable of
balancing  near a potential energy maximum. The common model of an energy
conserving point-contact gyro, however, does not have asymptotic stability
since it is Hamiltonian. Adding a rounded tip to the top, with nonholonomic
rolling contact, is not stabilizing.   A spinning top with dissipation,
however, can be asymptotically  stable in a transient sense in that, over a
limited time until the spinning rate has slowed too much, vertical motion is
approached exponentially. The observed asymptotic stability of rolling coins
and the like also depends on dissipation.

We know of only a few uncontrolled three-dimensional devices that can have
asymptotically stable steady motions at or near a potential energy maximum,
without depending on fast spinning parts. These devices are all
nonholonomically-constrained and conservative: (1) a no-hands bicycle with
massless wheels (say skates) and a special mass
distribution~\cite{hand:1988,papadopoulos:1988}; (2) closely related to the
bicycle is a rolling disk with eccentric masses that bank and steer but do not
pitch with the disk~\cite{papadopoulos:1988,coleman:1997}; (3) a no-hands
tricycle (where gyroscopic terms from the spinning wheels are not relevant for
balance because of the three point support) with a mildly soft de-centering
(negative spring constant) spring on the
steering~\cite{rocard:1960,sharp:1983};  (4) a rigid rider attached
appropriately to a moving skate-board~\cite{hubbard:1979};  and (5) a
statically unstable boat with an ideal keel, acting as a nonholonomic
constraint, that is steered by the boat lean similarly to how a bicycle front
wheel is steered by bike lean~\cite{cardanha:1991}.  Certain gliding aircraft
might also be considered as an example, but the definition of a potential
energy maximum is less clear for planes since there is no well defined
reference for measuring potential energy.

All of these devices differ from walking mechanisms in that they are
constrained against fore-aft tipping (the walking devices have fore-aft
dynamics), they conserve energy (the walkers lose energy at joint and foot
impacts and use up gravitational potential energy), and they are
nonholonomically constrained (most of the walkers are well modeled as
piecewise holonomic).  

\paragraph*{\bf Intermittent contact and nonholonomicity.} Mechanical systems
that are asymptotically stable must be non-Hamiltonian.   Two mechanisms for
losing the Hamiltonian structure of governing equations are dissipation and
nonholonomic constraints.  The primary examples of nonholonomic constraint are
rolling contact and skate-like sliding contact.  For these two smooth
constraints, and other less physical nonholonomic constraints, the 
set of allowed differential motions is not integrable. That is, the
constraints are not equivalent to a restriction of the space of
admissible configurations. For smooth nonholonomic systems, the dimension
of the configuration space accessible to the system is greater than the
dimension of the velocity space allowed by the  constraints.

An intermittent non-slipping contact constraint can also cause  the dimension
of the accessible configuration space to be greater than the dimension of the
accessible velocity space. As suggested by one simple
example~\cite{ruina:1997}, this discrete nonholonomicity may account for
exponential stability of some systems.   The walking models we study are all
nonholonomic in this intermittent sense (and also in the conventional sense if
they have rounded feet). They can, for example, translate forwards by walking
although the contact constraint does not allow forward sliding.   

\paragraph*{\bf Dynamical modeling.}

\begin{figure}[htp]
\centerline{\epsfxsize=3.5in\epsffile{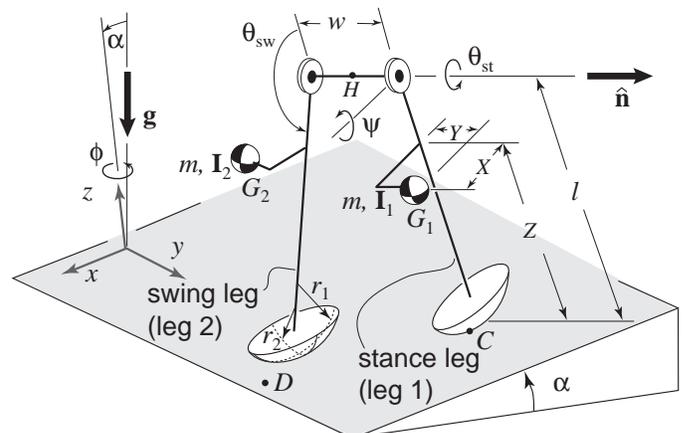}}
\caption{ A rigid body model of the simple walker. Parameters
and state variables are described in the text.}
\label{pointfoot3D.config}
\end{figure}

Fig.~\ref{pointfoot3D.config} shows a 3D model which probably captures  the
essential geometric and mass-distribution features of the physical model
presented here.  The device, at least at the level of approximation which we
believe is appropriate, is a pair of symmetric rigid bodies (leg 1 =
stance leg, leg 2 = swing leg) that have mass $m$, symmetrically located (in
the rest state) centers of mass $G_{1,2}$,  and mirror-symmetry related moment
of inertia matrices with respect to the center of mass ${\bf  I}_{1,2}$. The
legs are connected by a frictionless hinge at the hip with center point $H$ and
orientation $\hat{\bf  n}$ normal to the symmetry plane of the legs.  Each of
the two legs  can make rolling and collisional contact with the ground (slope
$=\alpha$)  with no contact couples.  The gravitational acceleration is
$\bf  g$.

The (reduced) dynamical state of the model is determined by the orientations
and angular velocities of the legs. The stance leg orientation is determined
by standard Euler angles $\psi$, $\theta_{\rm st}$, $\phi$ for lean, pitch and
steer. The configuration of the swing leg is described by the angle
$\theta_{\rm sw}$.  The absolute position of the walker on the plane does not
enter into the governing equations.  The instantaneous point of contact of the
stance leg with the ground is $C$ and the point of the impending contact is
$D$. We assume ground collisions are without bounce or slip.

The unreduced accessible configuration space is six-dimensional (the above
angles  plus position on the slope) whereas at any instant in time the
accessible velocity space is four-dimensional (the four dynamical state variables).
Hence the overall nonholonomicity ($6>4$) of this system which is smooth and 
holonomic at all but instants of collision. The model is also  dissipative due
to kinetic energy loss at the  collisions.

The model is well-posed  since the governing equations for rigid bodies in
hinged, rolling, and plastic-collisional contact are well established.    The
equations which govern the evolution of the state of the system ${\bf 
q}=\{\phi, \dot{\phi}, 
\psi,\dot{\psi}, \theta_{\rm st},\dot{\theta}_{\rm st}, 
\theta_{\rm sw}, \dot{\theta}_{\rm sw}\}$ follow from angular momentum balance
(or other equivalent principles).  Between collisions, we have angular
momentum balance for the whole system about the contact point $C$  
\begin{equation}
  \sum_{i=1,2}{\bf  r}_{_{\!G_i/C}}\! \times m{\bf  g}
  = \!\!
  \sum_{i=1,2} \left[ {\bf  r}_{_{\!G_i/C}}\! \times m{\bf  a}_i
   + 
\mbox{\boldmath$\omega$}_i \times 
\left({\bf  I}_i\mbox{\boldmath$\omega$}_i \right)
   + {\bf  I}_i \dot{\mbox{\boldmath$\omega$}}_i
\right]
 \label{goveqs1}
\end{equation}
where  ${\bf  r}_{G_i/C}\equiv {\bf  r}_{G_i}-{\bf  r}_C$, the center of
mass velocities and accelerations are ${\bf  v}_{1,2}$ and ${\bf  a}_{1,2}$,
and the angular velocities are
$\mbox{\boldmath$\omega$}_{1,2}$. Angular momentum balance for  the swing
leg about the hip axis $\hat{\bf  n}$ is
\begin{equation}
 \hat{\bf  n}\cdot  \left\{{\bf  r}_{_{\!G_2/H}} \times m{\bf  g}\right.
 =
    \left.{\bf  r}_{_{\!G_2/H}} \times m{\bf  a}_{_2}
     +\mbox{\boldmath$\omega$}_{_2} \times \left( {\bf 
I}_2\mbox{\boldmath$\omega$}_{_2} \right)
   + {\bf  I}_2 \dot{\mbox{\boldmath$\omega$}}_{_2}
\right\}
 \label{goveqs2}
\end{equation}
The  eight collisional jump conditions come from continuity of
configuration through the collision, conservation of angular momentum of the
system about the new contact point
$D$, 
\begin{equation}
 \left. \sum_{i=1,2}  {\bf  r}_{_{\!G_i/D}} \times m{\bf  v}_{_i}
   + {\bf  I}_i\mbox{\boldmath$\omega$}_i 
 \right|_{-}
   \!\!\!  = \!
  \left.
   \sum_{i=1,2}  {\bf  r}_{_{\!G_i/D}} \times m{\bf  v}_{_i}
   + {\bf  I}_i\mbox{\boldmath$\omega$}_{_i}   
   \right|_{+}
 \label{jump1}
\end{equation}
and conservation of angular momentum for the swing leg about the swing
hinge axis
\begin{equation}
    \hat{\bf  n}\cdot \!\left\{ \left.{\bf  r}_{_{\!G_1/H}} \times m{\bf 
v}_{_1}
   + {\bf  I}_1\mbox{\boldmath$\omega$}_{_1}  \right|_{-} \right.
      =
   \left. \left.{\bf  r}_{_{\!G_2/H}} \times m{\bf  v}_{_2}
   + {\bf  I}_2\mbox{\boldmath$\omega$}_{_2}  \right|_{+} \right\}
 \label{jump2}
\end{equation}
where the respective sides are to be evaluated just before ($-$) and after
($+$) foot collision with the ground.  The second jump condition 
Eqn.~(\ref{jump2}) is
being applied to the same leg as it switches from stance (subscript 1) to
swing (subscript 2).  Both jump conditions, 
Eqns.~(\ref{jump1})~and~(\ref{jump2}) also assume no collisional impulse from 
the ground to the leg which is just leaving the ground.

The governing equations and jump conditions above are expressed in terms of
positions, velocities, and accelerations, which are all complicated functions
of the state variables. As a result, the governing equations are massive
expressions (pages long).  We assembled the kinematic expressions and governing
differential equations using symbolic algebra software (MAPLE).

The no-slip rolling condition is that the velocity of
the material point in contact at $C$ is zero.  The acceleration of this point,
needed to calculate the accelerations of $G_{1,2}$,  is
given by $\mbox{\boldmath$\omega$}^{*}\cdot 
\mbox{\bf  R}\mbox{\boldmath$\omega$}^{*}$ where $\mbox{\boldmath$\omega$}^{*}$
is the in-the-contact-plane part of the angular velocity and {\bf  R}
is the inverse of the local surface curvature matrix. So far, we have only
studied a simplification with
point-contact feet ($r_{1} = r_{2} =0$, {\bf  R} is
the zero matrix) and no hip spacing ($w=0$).
In this case, when a foot is on the ground, the contact acts like a
ball-and-socket joint and the only nonholonomy is that of intermittent
contact. 

In order to study the stability of such systems, following McGeer, we
represent an entire gait cycle by a Poincar\'{e} map
\begin{equation}
{\bf  f}({\bf  q}_{k})={\bf  q}_{k+1}
\end{equation}
from the state of the system 
${\bf  q}_{k}$ {\em just} after a foot collision to the state 
${\bf  q}_{k+1}$ {\em just} after the next collision of the same foot (two leg 
swings and two foot collisions per map iteration).
We evaluate $\bf f$ using numerical integration
of Eqns.~(\ref{goveqs1})~and~(\ref{goveqs2}) between collisions and  applying 
the jump conditions Eqns.~(\ref{jump1})~and~(\ref{jump2}),  at each foot 
collision. For this model, the map is seven-dimensional ($8 - 1$), but we treat
it as eight-dimensional for numerical convenience.

Fixed points of the return map $\bf f$ 
($\bf q$ with ${\bf  f}({\bf q})={\bf  q}$) correspond to periodic gait
cycles (not necessarily stable).   We find
fixed points by numerical root finding on the function $\bf f-q$, sometimes
using fixed points from models with nearby parameter values to initialize
searches.
 
 We determine the stability of periodic motions by numerically
calculating the eigenvalues of the linearization of the return map at
the fixed points. If the magnitudes of some of the eigenvalues are less than
one (with all others equal to one), then the fixed point is asymptotically 
stable in those variables. Because there are a family of limit cycles at
different headings one eigenvalue is always one. Because we
use eight instead of seven dimensions in our map, one eigenvalue is always zero.

To date, like McGeer~\cite{mcgeer6:1992} and Fowble and
Kuo~\cite{kuo:1996} who studied similar simulations, we have found only
unstable periodic motions, though  less unstable than theirs.  A nearly stable
case from our numerical studies has maximum eigenvalue modulus of about 1.15,
one of exactly one, and the other six less than one. Fore-aft balance has 
already
been achieved with two-dimensional walking models whose stable fixed points we
use as starting points for the 3D analysis. Thus the eigenvector associated
with the maximum eigenvalue corresponds to falling over sideways (i. e., is
dominated by
$\psi,\dot{\psi}$ component) as  expected.  The most stable mass
distributions we have found  do not have very
human-like parameters; each leg  has a center of mass closer to the foot than
the hip, and laterally displaced at about $90\%$ of the leg length.

In this almost-stable case, the  walker's legs  have a mass distribution
corresponding roughly to laterally extended  balance bars, like what might be
used for walking on a tight-rope.   In the limit, as the lateral offset of the
center of mass gets very large, the device approaches, for sideways balance, an
inverted pendulum with large rotational inertia. The step periods remain
bounded. Negligible falling acceleration can thus occur in one step
and  the modulus of the maximum eigenvalue of the linearized step--to--step map
asymptotically approaches one, or apparent neutral stability, from {\em above}. 
Thus the closeness of the largest map eigenvalue modulus to one is not a 
complete measure of closeness to stability.  However, when averaged over a step cycle,
this model does fall more slowly than a corresponding inverted pendulum and
the low eigenvalue is not just a result of slowed falling due to large
rotary inertia.

\paragraph*{\bf The toy.}

\begin{figure}[htp]
\centerline{\epsfxsize=3.5in\epsffile{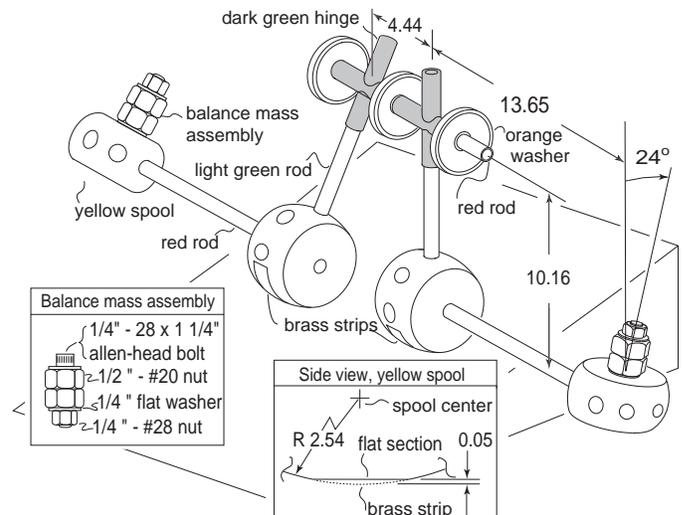}}
\caption{The 3D Tinkertoy$^{\circledR}$ walking
model with hardware description and dimensions (in centimeters,
not drawn to scale).
The balance masses and the brass strips
are fastened with black electrical tape (not shown).}
\label{tinker.toy}
\end{figure}

As a non-working demonstration of the kinematics and mass distributions in
our simulations, and not for walking experiments, we assembled a device similar
to the one shown in Fig.~\ref{tinker.toy}. It has two straight legs, separated
by  simple hinges at the hips, laterally extending balance mass rods, and
rounded feet. Playing, with no hopes of success, we
placed the toy on a ramp. Surprisingly, it  took a few serendipitous, if not
very steady or stable, steps.  After some non-quantifiable tinkering, we
arrived at the functioning device shown.

Our physical model is constructed from a popular American
child's construction toy,  brass strips to round
the feet bottoms, and various steel nuts for balance masses.  The walking ramp
has about a 4.5 degree slope and is narrow enough to  avoid making contact
with the balance masses as the walker rocks side-to-side.  Another more
complex assembly of similar toy parts (not described here) walks on a wide
ramp.

\paragraph*{\bf Aside: construction details.} 
The device
is built using the Playskool$^{\circledR}$ Tinkertoy$^{\circledR}$ Construction System:
Colossal Constructions$^{\mbox{TM}}$, 1991 set. One leg is made from a yellow
spool, a  light green rod, and a dark green hinge (plus `$+$' shaped) glued 
together. Then, we slid the legs onto a red rod (loose fit) which acts as an 
axle. The green hinges are separated and kept from sliding apart by three  
orange washers  friction-fit to the red axle. The legs and red axle can rotate 
independently.

To support the side weights, we glued a yellow spool rigidly to the end of a red
rod and inserted the other end into the side of a yellow foot 
with a friction fit to allow for rotational adjustment.

We assembled each balance mass from two stacked steel nuts held  together
between two washers by a nut and bolt. Each nut assembly has a mass of about 50
grams. Then, each balance mass assembly was located on the yellow spools at the
end of the balance rods and held in place with  vinyl electrical tape. 
The balance mass assembly  is tilted behind the leg.  As a result, the legs
have low mass centers located laterally at a distance comparable to the leg
length,  above the center of curvature of the feet, and just behind the leg
axes.  The mass of the fully assembled walking device is  about 120 grams,
only 20 grams more than the two balance masses.  When the toy is in  its
unstable-equilibrium  standing  position the nominally-vertical legs
are approximately orthogonal to the ramp.

Because a yellow spool has holes located radially around its
circumference to accept rods, a small flat section is on the bottom at
the foot contact point. To ensure that the walker is statically unstable
(cannot stand on the flat sections or in any other way), a small (0.50 cm wide)
strip of thin (0.013 cm)  brass shim  stock material was fastened over the flat
section contacting the floor so as to  restore its curvature there. 

\paragraph*{\bf Observed motion.}

Because the center of mass is above the center of curvature of the round
feet, we cannot stably stand this device with parallel or with splayed
legs. When placed aiming downhill on a ramp, tipped to one side, and released,
the device rocks side-to-side and, coupled
with swinging of the legs, takes tiny steps.  When a foot hits the
ground, it sticks and then rolls, until the swinging foot next collides with
the ground. 
Except at the moment of foot collision, only one foot is in contact with the
ground at any time.  When the swinging foot collides with the ground, the
trailing leg leaves the ground.  The gait is
more-or-less steady; after small disturbances the toy either falls or stumbles
a few steps while returning
to near-periodic gait.  At a slope of 4.5 degrees, it
takes a step about every 0.47 seconds and advances forward about 1.3 cm per
step, where a step is measured from a foot collision to the next collision of
that same foot. The side-to-side tilt is about 4 degrees, there is no visible
variation in $\phi$ during a step, but there is slight directional drift (one
way or another) over many steps. The rounded metal strips at the feet bottom
deform during foot collision  in a way that may or may not be essential; we do
not know.

\paragraph*{\bf Conclusions.}
We have constructed a device which can balance while walking but cannot stand
in any configuration. Although our new machine does not have a very human-like
mass distribution, it does highlight the possibility that uncontrolled
dynamics may not just contribute to  fore-aft walking balance, as indicated by
previous McGeer models, but also to side-to-side balance. The mechanism  
joins a small collection of statically unstable devices
which dynamically balance without any rapidly spinning parts.

Our too-simple mathematical/computational model does not explain this
behavior.  We do not yet know what key modeling features need be included
to predict the observed dynamic stability. An open and possibly unanswerable
question is whether the stability of this intermittently dissipative system
can be explained, in part,  by the fact that its piecewise holonomic
constraints act somewhat like  nonholonomic constraints.

\paragraph*{\bf Acknowledgments.}  Thanks to Les Schaffer, Saskya van Nouhuys,
and Mariano Garcia for editorial comments.


\end{document}